\documentclass[11pt]{article}
\usepackage{mathrsfs}
\usepackage{amssymb}
\usepackage{fullpage,latexsym}
\usepackage{graphicx}
\usepackage{epsfig}
\usepackage{psfrag}
\usepackage{subfigure}

\newtheorem{algorithm}{Algorithm}
\newtheorem{theorem}{Theorem}

\newtheorem{proposition}[theorem]{Proposition}
\newenvironment{proof}[1][Proof]{\noindent\textbf{#1.} }{\ \rule{0.5em}{0.5em}}

\newtheorem{definition}{Definition}

\newcommand{\qed}{\hfill \rule{2mm}{2mm}}
\newcommand{\pf}{{\bf Proof: }}
\newcommand{\comb}[2] {\mbox{$\left( { #1 \atop #2 } \right)$}}

\begin{document}
\title{Quantum algorithm to distinguish Boolean functions of
different weights}
\author{Samuel L. Braunstein, Byung-Soo Choi,\\
Department of Computer Science, University of York, York YO10 5DD,
United Kingdom.\\
Email: \{schmuel, bschoi\}@cs.york.ac.uk,
\and
Subhamoy Maitra, Dibyendu Chakrabarti, Subhroshekhar Ghosh,\\
Applied Statistics Unit, Indian Statistical Institute, Kolkata 700108, India.\\
Email: \{subho, dibyendu\_r, bst0219\}@isical.ac.in,
\and
Partha Mukhopadhyay\\
Institute of Mathematical Sciences, C. I. T Campus, Taramani,
Chennai 600 113, India.\\
Email: partha.mukherjee@rediffmail.com
}
\date{}
\maketitle

\begin{abstract}
By the weight of a Boolean function $f$, denoted by $wt(f)$, we mean
the number of inputs for which $f$ outputs $1$. Given a promise that
an $n$-variable Boolean function (available in the form of a black
box and the output is available in constant time once the input is
supplied) is of weight either $wN$ or $(1-w)N$ $(0 < w < 1, N=2^n)$,
we present a detailed study of quantum algorithms to find out which
one actually it is. To solve this problem we apply the Grover's
operator.

First we consider the restricted problem. Given a promise that an
$n$-variable Boolean function is of weight either
$\lfloor{N\sin^2{\frac{k}{2k+1}\frac{\pi}{2}}}\rceil$ or
$\lfloor{N\cos^2{\frac{k}{2k+1}\frac{\pi}{2}}}\rceil$
($\lfloor{q}\rceil$ means the nearest integer corresponding to the
real value $q$), we show that one can suitably apply Grover's
operator for $k$-many iterations to decide which case this is with a
probability almost unity for large $n$ and $k$ in $O(poly(n))$. On
the other hand, the best known probabilistic classical algorithm has
a success probability close to $0.5$ (from above) after $k$ many
steps when $k$ is large. We further show that the best known
probabilistic classical algorithm can achieve a success probability
almost unity only after $k^s$ many iterations where $s > 2$. This
indicates a quadratic speed up (and also agrees to the quadratic
speed up by the use of Grover's algorithm in database search) on
time complexity in the quantum domain with respect to the best known
result in the classical domain.

Second, we modify the basic randomized algorithm into a sure success
algorithm, which can distinguish Boolean functions of weights $wN$
or $(1-w)N$ for any $w$, $(0 < w < 1)$. To do that we have exploited
a sure success Grover search algorithm, which modifies the very last
operation. For the weight decision problem, we show that the very
last two operations should be changed to distinguish any weight with
certainty and found the phase conditions for the last two
operations.

As quantum counting methods exist, which can count the
number of solutions, here we compare our method with that.
Since the quantum counting method needs to exploit
period information, which requires many Grover operations, we have
found that our method is faster than the quantum counting method.
\end{abstract}

{\bf Keywords:} Quantum Algorithm, Boolean Function, Grover's Operator,
Weight Decision Problem.

\section{Introduction}
David Deutsch designed a quantum algorithm, which evaluates whether
the two outputs of a Boolean function is the same or not using only
one function evaluation \cite{CS-0900.81019MA}. Deutsch-Jozsa
generalized the Deutsch's algorithm for more general case such as
whether the Boolean function is constant or balanced \cite{qDJ92}.
Deutsch-Jozsa algorithm had been proposed to show an exponential speed-up
in quantum machine than the classical machines. The most important
contribution in this area has been achieved when Peter Shor discovered
a polynomial-time quantum algorithm for factoring and computing
discrete logarithms, which is exponentially faster than the
classical best known algorithm \cite{SICOMP::Shor1997}. Because this
quantum factoring algorithm works in polynomial time compared to
exponential time in classical method, many researchers started to
find other applications. Lov Grover discovered a quantum database
search algorithm, which is quadratic faster than classical database
search algorithm \cite{qGR96}. Meanwhile, database search algorithm
is one of the most widely used algorithm in the computer
applications, the impact is so huge and more researchers are
interested in other applications of quantum database search
algorithm. This research also focuses on the applications of Grover
database search algorithm.

In this research, we assume that there is a promise in the Boolean
function as it has a weight $wN$ or $(1-w)N$. Initially, for better
understanding, we formulate the problem for some special weight
cases such as $\lfloor N\sin^2(\frac{k}{2k+1}\frac{\pi}{2}) \rceil$
or $\lfloor N\cos^2(\frac{k}{2k+1}\frac{\pi}{2}) \rceil$, by
applying $k$ many Grover operations. Here the quantum algorithm
presented is of randomized in nature, but the success probability is
arbitrarily close to unity. Next we consider the general case to
distinguish functions having weight either $wN$ or $(1-w)N$ and we
also consider a deterministic algorithm. This requires changes in
the very last two Grover operations with phase conditions. Briefly,
from $1^{st}$ to $(k-2)^{th}$ steps, the original Grover operators
are used, but, in the last two steps, two different Grover operators
are used with the phase conditions. Also we found that the phase
conditions depending on the required number of Grover operations.
Meanwhile, because quantum counting algorithm was already proposed,
we compared two methods. Since the quantum counting method requires
more Grover iterations to find period information, we can conclude
that our method is faster than this method.

\section{Preliminaries}

A Boolean function $f(x_1, \ldots, x_n)$ on $n$ variables may be
viewed as a mapping from $\{0, 1\}^n$ into $\{0, 1\}$. A Boolean
function $f$ is constant if $f(x) = c$ for all $x \in \{0, 1\}^n$,
$c \in \{0, 1\}$. That means $wt(f)$ is either $0$ or $N$. A Boolean
function $f$ is balanced if $wt(f) = N/2$.

Given a promise that the function $f$ is either constant or
balanced, one may ask for an algorithm, that can exactly answer
which case it is. Note that throughout this document we consider
that any Boolean function $f$ is available in the form of an oracle
(black box) only, where one can apply an input to the black box to
get the output. A classical algorithm needs to check the function
for $N/2 + 1$ inputs in the worst case to decide whether the
function is constant or balanced.
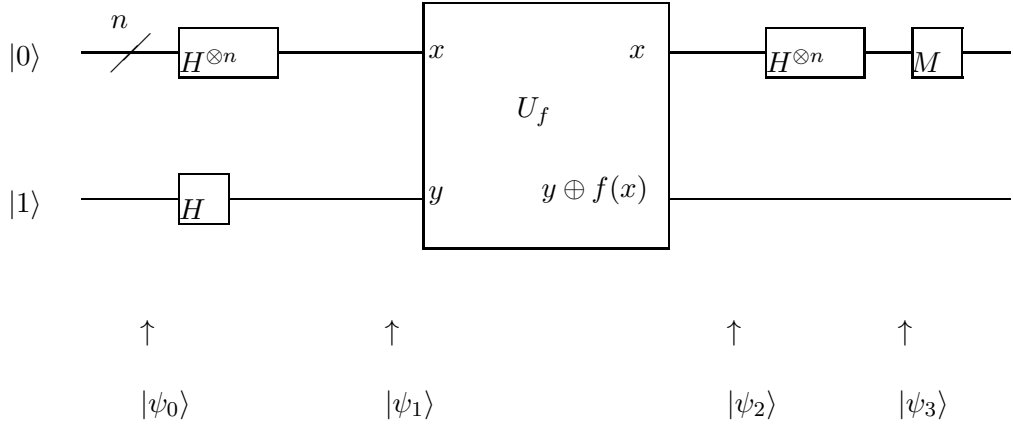
\begin{figure}[ht]
\setlength{\unitlength}{6.5mm}
\begin{picture}(25,9)
\put(1.3,7.6){\makebox(0,0)[bl]{ {$|0\rangle$}}}
\put(1.3,4.6){\makebox(0,0)[bl]{ {$|1\rangle$}}}
\put(10,4){\framebox(5,5)}
\put(3,5){\line(1,0){2}}
\put(3.6,7.6){\line(1,1){0.8}}
\put(3.4,8.5){\makebox(0,0)[bl]{ {$n$}}}
\put(5,4.5){\framebox(1,1)}
\put(4.8,4.6){\makebox(0,0)[bl]{ {$H$}}}
\put(6,5){\line(1,0){4}}
\put(15,5){\line(1,0){7}}
\put(3,8){\line(1,0){2}}
\put(5,7.5){\framebox(2,1)}
\put(4.8,7.6){\makebox(0,0)[bl]{ {$H^{\otimes n}$}}}
\put(7,8){\line(1,0){3}}
\put(15,8){\line(1,0){2}}
\put(17,7.5){\framebox(2,1)}
\put(16.8,7.6){\makebox(0,0)[bl]{ {$H^{\otimes n}$}}}
\put(19,8){\line(1,0){1}}
\put(20,7.5){\framebox(1,1)}
\put(19.75,7.6){\makebox(0,0)[bl]{ {$M$}}}
\put(21,8){\line(1,0){1}}

\put(9.9,4.9){\makebox(0,0)[bl]{ {$y$}}}
\put(9.9,7.9){\makebox(0,0)[bl]{ {$x$}}}
\put(14,7.9){\makebox(0,0)[bl]{ {$x$}}}
\put(12.2,4.9){\makebox(0,0)[bl]{ {$y \oplus f(x)$}}}
\put(11.7,6.5){\makebox(0,0)[bl]{ {$U_f$}}}

\put(4,2){\makebox(0,0)[bl]{{ $\uparrow$}}}
\put(9,2){\makebox(0,0)[bl]{{ $\uparrow$}}}
\put(16,2){\makebox(0,0)[bl]{{ $\uparrow$}}}
\put(19.5,2){\makebox(0,0)[bl]{{ $\uparrow$}}}
\put(4,0.5){\makebox(0,0)[bl]{{ $| \psi_0 \rangle$}}}
\put(9,0.5){\makebox(0,0)[bl]{{ $| \psi_1 \rangle$}}}
\put(16,0.5){\makebox(0,0)[bl]{{ $| \psi_2 \rangle$}}}
\put(19.5,0.5){\makebox(0,0)[bl]{{ $| \psi_3 \rangle$}}}
\end{picture}
\caption{Quantum circuit for Deutsch-Jozsa Algorithm}
\label{fig1}
\end{figure}
It is known that given a classical circuit for $f$, there is a
quantum circuit of comparable efficiency which performs a transformation
$U_f$ that takes input like $|x, y\rangle$ and produces the output
$|x, y \oplus f(x)\rangle$. Given such a $U_f$, Deutsch-Jozsa~\cite{qDJ92}
provided a quantum algorithm that can solve this problem in constant time,
indeed, in a single evaluation of $U_f$. The circuit for their algorithm
is given in Fig.~\ref{fig1}.

\begin{algorithm} {\em Deutsch-Jozsa Algorithm \cite{qDJ92}}
\label{algo:deutsch_jozsa_algorithm}
\begin{center}
\begin{tabular}{|ll|}
\hline
1. & $|\psi_0\rangle = |0\rangle^{\otimes n} |1\rangle$\\
2. & $|\psi_1\rangle = \frac{1}{\sqrt{2N}}\sum_{x \in \{0, 1\}^n}
|x\rangle(|0\rangle - |1\rangle)$\\
3. & $|\psi_2\rangle = \frac{1}{\sqrt{2N}}\sum_{x \in \{0, 1\}^n}
(-1)^{f(x)}|x\rangle (|0\rangle - |1\rangle)$\\
4. & $|\psi_3\rangle = \frac{1}{N\sqrt{2}}\sum_{x,z \in \{0, 1\}^n}
(-1)^{x z \oplus f(x)}|z\rangle (|0\rangle - |1\rangle)$\\
5. & {\rm Measurement at $M$: all-zero state implies that the} \\
   & {\rm function is constant, otherwise it is balanced.}\\
\hline
\end{tabular}
\end{center}
\end{algorithm}

The Deutsch-Jozsa algorithm yields an exponential speed-up relative to
any exact classical computation. This provides a relativized separation
between EQP and P with respect to the oracle $f$ (see~\cite{qNC02} for
basic notions of complexity theory).

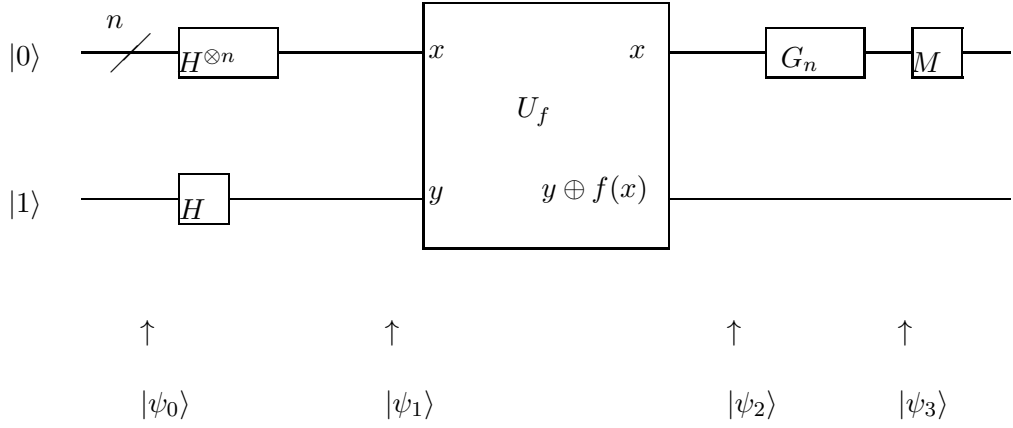
\begin{figure}[ht]
\setlength{\unitlength}{6.5mm}
\begin{picture}(25,9)
\put(1.3,7.6){\makebox(0,0)[bl]{ {$|0\rangle$}}}
\put(1.3,4.6){\makebox(0,0)[bl]{ {$|1\rangle$}}}
\put(10,4){\framebox(5,5)}
\put(3,5){\line(1,0){2}}
\put(3.6,7.6){\line(1,1){0.8}}
\put(3.3,8.5){\makebox(0,0)[bl]{ {$n$}}}
\put(5,4.5){\framebox(1,1)}
\put(4.8,4.6){\makebox(0,0)[bl]{ {$H$}}}
\put(6,5){\line(1,0){4}}
\put(15,5){\line(1,0){7}}
\put(3,8){\line(1,0){2}}
\put(5,7.5){\framebox(2,1)}
\put(4.75,7.6){\makebox(0,0)[bl]{ {$H^{\otimes n}$}}}
\put(7,8){\line(1,0){3}}
\put(15,8){\line(1,0){2}}
\put(17,7.5){\framebox(2,1)}
\put(17.1,7.63){\makebox(0,0)[bl]{ {$G_n$}}}
\put(19,8){\line(1,0){1}}
\put(20,7.5){\framebox(1,1)}
\put(19.75,7.6){\makebox(0,0)[bl]{ {$M$}}}
\put(21,8){\line(1,0){1}}

\put(9.9,4.9){\makebox(0,0)[bl]{ {$y$}}}
\put(9.9,7.9){\makebox(0,0)[bl]{ {$x$}}}
\put(14,7.9){\makebox(0,0)[bl]{ {$x$}}}
\put(12.2,4.9){\makebox(0,0)[bl]{ {$y \oplus f(x)$}}}
\put(11.7,6.5){\makebox(0,0)[bl]{ {$U_f$}}}

\put(4,2){\makebox(0,0)[bl]{{ $\uparrow$}}}
\put(9,2){\makebox(0,0)[bl]{{ $\uparrow$}}}
\put(16,2){\makebox(0,0)[bl]{{ $\uparrow$}}}
\put(19.5,2){\makebox(0,0)[bl]{{ $\uparrow$}}}
\put(4,0.5){\makebox(0,0)[bl]{{ $| \psi_0 \rangle$}}}
\put(9,0.5){\makebox(0,0)[bl]{{ $| \psi_1 \rangle$}}}
\put(16,0.5){\makebox(0,0)[bl]{{ $| \psi_2 \rangle$}}}
\put(19.5,0.5){\makebox(0,0)[bl]{{ $| \psi_3 \rangle$}}}
\end{picture}
\caption{Quantum circuit to distinguish $wt(f) = N/4$ and $wt(f) =
3N/4$} \label{fig2}
\end{figure}

Now we discuss a constant time quantum algorithm to distinguish
Boolean functions of weight $N/4$ and $3N/4$~\cite{GP01}. We replace
$H^{\otimes n}$ by Grover's matrix at the output side of $U_f$ in
Fig.~\ref{fig1} to get a circuit shown in Fig.~\ref{fig2} and show
that this solves the problem. In 2001, Green and Pruim~\cite{GP01}
presented a relativized separation between BQP and ${\mbox
P}^{\mbox{NP}}$ using a nice technique based on Grover's
algorithm~\cite{qGR96}. Green and Pruim's work relied on a
complexity theoretic formulation, whereas our analysis here is
directly related to weights of Boolean functions. Note that a
similar question has been discussed in~\cite[Section 5]{BO98}. There
also the problem was not exactly posed as a discrimination problem,
but as a search problem.

We denote the $N \times N$ Grover's matrix $G_n$ as
$G_n = H^{\otimes n}(2|0\rangle\langle 0|-{\bf{1}}) H^{\otimes n}
= \frac{2}{N} \sum_{x,y}|x\rangle\langle y|-{\bf{1}}$.
It is known that this operation may be constructed with $O(\log N)$
quantum gates~\cite{qNC02}. The circuit is shown in Fig.~\ref{fig2} and the
steps of the algorithm are as follows.

\begin{algorithm} {\em Randomized Algorithm to distinguish $wt(f) = 3N/4$ 
or $N/4$}
\label{algo:sure_success_quarter_decision}
\begin{center}
\begin{tabular}{|ll|}
\hline
1. & $|\psi_0\rangle = |0\rangle^{\otimes n} |1\rangle$\\
2. & $|\psi_1\rangle = \sum_{x \in \{0, 1\}^n}
\frac{|x\rangle}{\sqrt{N}} \frac{(|0\rangle - |1\rangle)} {\sqrt{2}}$\\
3. & $|\psi_2\rangle = \sum_{x \in \{0, 1\}^n}
\frac{(-1)^{f(x)}|x\rangle}{\sqrt{N}}\frac{(|0\rangle - |1\rangle)}
{\sqrt{2}}$. \\
   & {\rm Let} $|\psi^{\prime}_2\rangle = \sum_{x \in \{0, 1\}^n}
     \frac{(-1)^{f(x)}|x\rangle}{\sqrt{N}}$.\\
4. & $|\psi_3\rangle = G_n |\psi^{\prime}_2\rangle
\frac{(|0\rangle - |1\rangle)}{\sqrt{2}}$\\
5. & {\rm Measure the resulting state} $G_n |\psi^{\prime}_2\rangle$
     {\rm in the} \\
   & {\rm computational basis and let the result be ${\hat x}$.}\\
6. & {\rm if $f({\hat x})$ = 0 then $wt(f) = 3N/4$
else $wt(f) = N/4$.}\\
\hline
\end{tabular}
\end{center}
\end{algorithm}
The key result proving the correctness of this algorithm is as follows and we
present a proof as it will be discussed in details in the following section.
\begin{theorem}
\label{qth1} $G_n |\psi^{\prime}_2\rangle = \sum_{x : f(x) = 0}
\frac{N-4 \,wt(f)}{N\sqrt{N}} |x\rangle + \sum_{x : f(x) = 1}
\frac{3N-4\, wt(f)}{N\sqrt{N}} |x\rangle$, {\rm and
Algorithm~\ref{algo:sure_success_quarter_decision} produces a
correct result.}
\end{theorem}
\pf $G_n |\psi^{\prime}_2\rangle\ = \frac{1}{\sqrt{N}} G_n
\sum_{x\in \{0,1\}^n} (-1)^{f(x)}|x\rangle =
\frac{1}{\sqrt{N}}\Bigl( \frac{4N-2 wt(f)}{N}  \sum_x |x\rangle -
\sum_x (-1)^{f(x)}|x\rangle\Bigr)$. From which the result follows.

If $wt(f) = N/4$ then the probability amplitude of all the
$|x\rangle$ for which $f(x) = 0$ vanishes. So on measurement we will
get some ${\hat x}$ for which $f({\hat x})$ is 1.  On the other hand,
if $wt(f) = 3N/4$ then the probability amplitude of all the
$|x\rangle$ for which $f(x) = 1$ vanishes. So on measurement we will
get some ${\hat x}$ for which $f({\hat x})$ is 0. \qed

\section{Repeated Application of Grover's Operator}
\label{Repeat} Note that Grover's search algorithm~\cite{qGR96} uses
repeated applications of $G_n$. Motivated by the same idea we now
analyze in detail the repeated application of $G_n U_f$. One may
refer the important papers
like~\cite{2000quant.ph..5055B,1998quant.ph..5082B,BO98,HOYER,LI-HWANG-HSIEH-WANG,HSIEH-LI,LONG-TU-LI-ZHANG-YAN,LONG,MM01,AN98,LONG-LI-SUN}
where the Grover's operator has been used. We also make an elaborate
study to present a comprehensive understanding of this problem in
this section. We start with a modification of
Algorithm~\ref{algo:sure_success_quarter_decision}.
\begin{algorithm} {\em Randomized Algorithm for Weight Decision Problem
$3N/4$ and $N/4$}
\label{algo:randomized_weight_decision_algorithm}
\begin{center}
\begin{tabular}{|ll|}
\hline
1. & $|\psi_0\rangle = |0\rangle^{\otimes n} |1\rangle$\\
2. & {\rm Let us denote} $|\psi^{\prime}_1\rangle =
\sum_{x \in \{0, 1\}^n} \frac{|x\rangle}{\sqrt{N}}$\\
3. & $i = 0$.\\
4. & $|\psi_2\rangle = U_f(|\psi^{\prime}_1\rangle
\frac{(|0\rangle - |1\rangle)} {\sqrt{2}})$\\
5. & {\rm Let us denote first $n$ qubits of
    $|\psi_2\rangle$ as $|\psi^{\prime}_2\rangle$.}\\
6. & $|\psi_2\rangle = |\psi^{\prime}_2\rangle
        \frac{(|0\rangle - |1\rangle)}{\sqrt{2}}$. \\
7. & $|\psi_3\rangle = G_n |\psi^{\prime}_2\rangle
\frac{(|0\rangle - |1\rangle)}{\sqrt{2}}$.
{\rm Let} $|\psi_3^{\prime}\rangle = G_n |\psi^{\prime}_2\rangle$.\\
8. & $i = i+1$.\\
9. & {\rm If $(i < k)$ denote $|\psi^{\prime}_3\rangle$ by
$|\psi^{\prime}_1\rangle$ and go to step 4.}\\
10. & {\rm Measure the resulting state} $G_n |\psi^{\prime}_2\rangle$
     {\rm in the} \\
   & {\rm computational basis and let the result be ${\hat x}$.}\\
11-1. & {\rm if $k$ is odd and if $f({\hat x})$ = 0 then $wt(f) =
\lfloor{N\cos^2{\frac{k}{2k+1}\frac{\pi}{2}}}\rceil$}\\
  & {\rm else
$wt(f) = \lfloor{N\sin^2{\frac{k}{2k+1}\frac{\pi}{2}}}\rceil$.}\\
11-2. & {\rm if $k$ is even and if $f({\hat x}) = 1$ then $wt(f) =
\lfloor{N\cos^2{\frac{k}{2k+1}\frac{\pi}{2}}}\rceil$}\\
  & {\rm else
$wt(f) = \lfloor{N\sin^2{\frac{k}{2k+1}\frac{\pi}{2}}}\rceil$.}\\
\hline
\end{tabular}
\end{center}
\end{algorithm}
In this section we will consider a number of iterations $k \geq 1$
and then show how we can distinguish the weights with a very high
(almost unity) success probability. Without loss of generality we
detail the analysis by taking odd $k$.

\begin{theorem}
\label{thaa}
Let $a_k$ denotes the amplitude of the states $|x\rangle$
where $f(x) = 0$ and $b_k$ denotes the amplitude of the state $|x\rangle$
where $f(x)= 1$ after $k$-th iteration. Then,
$a_k = (2 \frac{N - wt(f)}{N} - 1) a_{k-1} - 2 \frac{wt(f)}{N} b_{k-1}$,
$b_k = 2\frac{N - wt(f)}{N} a_{k-1} + (2 \frac{N - wt(f)}{N} -1) b_{k-1}$,
with initial conditions $a_0 = b_0 = \frac{1}{\sqrt{N}}$.
\end{theorem}
\pf The proof follows from
Algorithm~\ref{algo:randomized_weight_decision_algorithm} in a
similar fashion as the proof of Theorem~\ref{qth1}. \qed

Our interest is to investigate the zeros of $a_k$
and $b_k$. It may be noted that the solutions to the recurrence relations are
given by
$$a_{k} = \frac{1}{\sqrt{N}}  \frac{\sin{(2k + 1)\beta}}{\sin{\beta}},
b_{k} = \frac{1}{\sqrt{N}}  \frac{\cos{(2k +
1)\beta}}{\cos{\beta}},$$ where  $\sin^2{\beta} = \frac{wt(f)}{N}$
($= u$, say for notational convenience). Note that this recursion
and Theorem~\ref{thaa} have been described in~\cite{BO98}. Clearly,
the factor $1/\sqrt{N}$ does not play any part in determining the
zeros of $a_{k}$ and $b_{k}$. The zeros of $a_{k}$ and $b_{k}$ are
given by
\[\sin(2k + 1)\beta  = 0, \sin{\beta} \neq 0\
\mbox{ and } \cos(2k + 1)\beta = 0, \cos{\beta} \neq 0\]
respectively. Now $\sin{(2k + 1)\beta} = 0 \Rightarrow (2k + 1)\beta
= l \pi \Rightarrow \beta = \frac{l\pi}{(2k + 1)}$, where $l \in
\mathbb{Z}$. Also $\cos{(2k + 1)\beta} = 0 \Rightarrow (2k + 1)\beta
= (2m - 1)\frac{\pi}{2} \Rightarrow \beta = \frac{(2m - 1)}{(2k +
1)}\frac{\pi}{2}$ where $m \in \mathbb{Z}$.

As we are interested in distinct roots $u = \sin^2{\beta}$ of $a_k =
0$ (respectively $b_k = 0$), it is clear that we will get the
distinct roots when $1 \leq l \leq k$ (respectively $1 \leq m \leq
k$). We can summarize the above discussion in the following result.

\begin{proposition}
\label{prop1} The $k$ distinct roots of $a_k = 0$ and $b_k = 0$ are
$\sin^{2}{\frac{l\pi}{(2k + 1)}}$ and $\sin^{2}{\frac{(2m -
1)}{(2k+1)}\frac{\pi}{2}}$ respectively where $1 \leq l,m \leq k$.
\end{proposition}

\begin{proposition}
For each root of the equation $a_{k} = 0$ there is a corresponding root of
the equation $b_{k} = 0$ so that their sum is 1.
\end{proposition}
\pf The roots of $a_{k} = 0$ and $b_k = 0$ are of the forms
$\sin^{2}{\frac{l\pi}{(2k + 1)}}$ and $\sin^{2}{\frac{(2m -
1)}{(2k+1)}\frac{\pi}{2}}$ respectively where $1 \leq l,m \leq k$.
Let us consider the pairings of $l,m$ such that $l + m = k+1$. Now $
\sin^{2}{\frac{l\pi}{(2k + 1)}} + \sin^{2}{\frac{(2m -
1)}{(2k+1)}\frac{\pi}{2}} = 1$, which gives the proof. \qed

Given any $k$,
Algorithm~\ref{algo:randomized_weight_decision_algorithm} can
distinguish whether a Boolean function is either from the weights
$\lfloor{N\sin^{2}{\frac{(2m - 1)}{(2k+1)}\frac{\pi}{2}}}\rceil$ or
from the weights $\lfloor{N\cos^{2}{\frac{(2m -
1)}{(2k+1)}\frac{\pi}{2}}}\rceil$, with good success probability for
$m = 1, \ldots, k$. However, here we are interested in
distinguishing two Boolean functions which are closest in weight to
balanced functions, i.e., of weight $\frac{N}{2} = 2^{n-1}$.
\begin{definition}
Let $r_{a, k}$ be the root of $a_k = 0$ such that
$|r_{a, k} - 0.5| \leq |\rho_{a, k} - 0.5|$ for any root $\rho_{a, k}$
of $a_k = 0$. Similarly let $r_{b, k}$ be the root of $b_k = 0$ such that
$|r_{b, k} - 0.5| \leq |\rho_{b, k} - 0.5|$ for any root $\rho_{b, k}$
of $b_k = 0$. Let us denote $\mu_k = \min\{r_{a, k}, r_{b, k}\}$.
\end{definition}
\begin{proposition}
\label{prop3}
\begin{eqnarray}
\mu_k &=& r_{a, k} = \sin^2{\frac{2\alpha-1}{4\alpha-1}\frac{\pi}{2}},
\mbox{ when } k = 2\alpha - 1 \mbox{ and } m = \alpha \nonumber \\
\     &=& r_{b, k} = \sin^2{\frac{2\alpha}{4\alpha+1}\frac{\pi}{2}},
\mbox{ when } k = 2\alpha \mbox{ and } l = \alpha. \nonumber
\end{eqnarray}
\end{proposition}
\pf Let $k = 2\alpha - 1$. For $m = \alpha$, the root of $a_k = 0$ is
$\rho_{a, k} = \sin^2{\frac{2\alpha-1}{4\alpha-1}\frac{\pi}{2}}$.
For $m = \alpha + 1$, the root of $a_k = 0$ is
$\rho^\prime_{a, k} = \sin^2{\frac{2\alpha+1}{4\alpha-1}\frac{\pi}{2}}$.
As $2\alpha + 1 > 2\alpha$,
$\sin^2{\frac{2\alpha+1}{4\alpha-1}\frac{\pi}{2}} >
\sin^2{\frac{2\alpha}{4\alpha-1}\frac{\pi}{2}}$, which gives,
$\sin^2{\frac{2\alpha+1}{4\alpha-1}\frac{\pi}{2}} +
\cos^2{\frac{2\alpha}{4\alpha-1}\frac{\pi}{2}} > 1$, i.e.,
$\sin^2{\frac{2\alpha+1}{4\alpha-1}\frac{\pi}{2}} +
\sin^2{\frac{2\alpha-1}{4\alpha-1}\frac{\pi}{2}} > 1$, i.e.,
$\sin^2{\frac{2\alpha+1}{4\alpha-1}\frac{\pi}{2}} - \frac{1}{2} >
\frac{1}{2} - \sin^2{\frac{2\alpha-1}{4\alpha-1}\frac{\pi}{2}}$ which gives
$\rho^\prime_{a, k} - \frac{1}{2} > \frac{1}{2} - \rho_{a, k}$.
Since all the other roots of $a_k = 0$ are either less than
$\rho_{a, k}$ or greater than $\rho^\prime_{a, k}$, we get
$r_{a, k} = \rho_{a, k}$. It is also clear that $r_{b, k} = 1 - r_{a, k}$
and hence here $r_{b, k} > r_{a, k}$. So $\mu_k = r_{a, k}$.

Let $k = 2\alpha$. For $m = \alpha$, the root of $b_k = 0$ is
$\rho_{b, k} = \sin^2{\frac{2\alpha}{4\alpha+1}\frac{\pi}{2}}$.
For $m = \alpha + 1$, the root of $b_k = 0$ is
$\rho^\prime_{a, k} = \sin^2{\frac{2\alpha+2}{4\alpha+1}\frac{\pi}{2}}$.
As $2\alpha + 2 > 2\alpha + 1$,
$\sin^2{\frac{2\alpha+2}{4\alpha+1}\frac{\pi}{2}} >
\sin^2{\frac{2\alpha+1}{4\alpha+1}\frac{\pi}{2}}$, which gives,
$\sin^2{\frac{2\alpha+2}{4\alpha+1}\frac{\pi}{2}} +
\cos^2{\frac{2\alpha+1}{4\alpha+1}\frac{\pi}{2}} > 1$, i.e.,
$\sin^2{\frac{2\alpha+2}{4\alpha+1}\frac{\pi}{2}} +
\sin^2{\frac{2\alpha}{4\alpha+1}\frac{\pi}{2}} > 1$, i.e.,
$\sin^2{\frac{2\alpha+2}{4\alpha+1}\frac{\pi}{2}} - \frac{1}{2} >
\frac{1}{2} - \sin^2{\frac{2\alpha}{4\alpha+1}\frac{\pi}{2}}$ which gives
$\rho^\prime_{b, k} - \frac{1}{2} > \frac{1}{2} - \rho_{b, k}$.
Since all the other roots of $b_k = 0$ are either less than
$\rho_{b, k}$ or greater than $\rho^\prime_{b, k}$, we get
$r_{b, k} = \rho_{b, k}$. It is also clear that $r_{a, k} = 1 - r_{b, k}$
and hence here $r_{a, k} > r_{b, k}$. So $\mu_k = r_{b, k}$. \qed

\begin{theorem}
\label{th1}
$\mu_k = \sin^2{\frac{k}{2k+1}\frac{\pi}{2}}$ and $\mu_k < \mu_{k+1} < 0.5$.
\end{theorem}
\pf From Proposition~\ref{prop3}, it is clear that
$\mu_k = \sin^2{\frac{k}{2k+1}\frac{\pi}{2}}$. So
$\mu_{k+1} = \sin^2{\frac{k+1}{2k+3}\frac{\pi}{2}}$.
Now $\frac{k+1}{2k+3} - \frac{k}{2k+1} = \frac{1}{(2k+3)(2k+1)} > 0$,
which gives $\mu_{k+1} > \mu_k$. Further it is easy to see that $\mu_{k+1} =
\sin^2{\frac{k+1}{2k+3}\frac{\pi}{2}} < \sin^2{\frac{\pi}{4}} <
\frac{1}{2}$. \qed

As $a_k$ and $b_k$ can be seen as polynomials in $u$, we now refer
them as $a_k(u)$ and $b_k(u)$, respectively. It is clear that
$a_k(\mu_k) = 0$. Now using
Algorithm~\ref{algo:randomized_weight_decision_algorithm}, we can
distinguish two Boolean functions of weight $\mu_k N$ and
$(1-\mu_k)N$. Unfortunately, $\mu_k N$ may not be an integer and in
that case we have to consider a Boolean function of weight
$\mu_k^\prime N$, where $\mu_k^\prime N$ is an integer and
$|\mu_k^\prime N - \mu_k N| \leq 0.5$. Thus we will be using the
Algorithm~\ref{algo:randomized_weight_decision_algorithm} to
distinguish between Boolean functions of weight $\mu_k^\prime N$ and
$(1-\mu_k^\prime)N$. This will incorporate some error in the
decision process. However, we will show that this error is almost
zero for large $N$.

\begin{theorem}
\label{mainthm} Consider Boolean functions on $n$ variables and let
$N = 2^n$. After $k$ iterations, $k$ in $O(poly(n))$, the quantum
algorithm
(Algorithm~\ref{algo:randomized_weight_decision_algorithm}) can
distinguish two Boolean functions of weights
$\lfloor{N\sin^2{\frac{k}{2k+1}\frac{\pi}{2}}}\rceil$ and
$\lfloor{N\cos^2{\frac{k}{2k+1}\frac{\pi}{2}}}\rceil$ with success
probability $> 1-\frac{64(k+1)^2}{N^2}$ which is almost unity for
large $N$.
\end{theorem}
\pf We have $a_{k} = \frac{1}{\sqrt{N}}  \frac{\sin{(2k +
1)\beta}}{\sin{\beta}} = 0$, when $\sin^2{\beta} = \mu_k$. Let
$\mu_k^\prime = \sin^2{\beta^\prime}$. We like to calculate the
value of $a_k(\mu_k^\prime) = \frac{1}{\sqrt{N}} \frac{\sin{(2k +
1)\beta^\prime}}{\sin{\beta^\prime}}$.

As $|\mu_k^\prime N - \mu_k N| \leq 0.5$, we get
$|\sin^2{\beta^\prime} - \sin^2{\beta}| \leq \frac{0.5}{N}$. Thus
$(\sin{\beta^\prime} + \sin{\beta}) |\sin{\beta^\prime} -
\sin{\beta}| \leq \frac{0.5}{N}$, i.e.,
$(2\sin{\frac{\beta^\prime+\beta}{2}}
\cos{\frac{\beta^\prime-\beta}{2}})
(2\cos{\frac{\beta^\prime+\beta}{2}}
|\sin{\frac{\beta^\prime-\beta}{2}}|) \leq \frac{0.5}{N}$, i.e.,
$(2\sin{\frac{\beta^\prime+\beta}{2}}
\cos{\frac{\beta^\prime+\beta}{2}})
(2\cos{\frac{\beta^\prime-\beta}{2}}
|\sin{\frac{\beta^\prime-\beta}{2}}|) \leq \frac{0.5}{N}$, i.e.,
$\sin{(\beta^\prime+\beta)}|\sin{(\beta^\prime-\beta)}| \leq
\frac{0.5}{N}$. This implies, $|\sin{(\beta^\prime-\beta)}| \leq
\frac{1}{2N \sin{(\beta^\prime+\beta)}}$. Note that $\frac{1}{4} =
\mu_1 < \mu_k < \frac{1}{2}$ for $k > 1$. Now $\mu_k =
\sin^2{\beta}$. So, $\frac{\pi}{6} < \beta < \frac{\pi}{4}$. As
$\beta \approx \beta^\prime$, $\beta + \beta^\prime \approx 2\beta$.
Due to the small difference between $\beta$ and $\beta^\prime$, it
may happen that on the lower side $\beta + \beta^\prime$ may
marginally be less than $2\frac{\pi}{6}$ and at the higher side may
marginally exceed $2\frac{\pi}{4}$. Thus it is safe to assume
$\sin{(\beta+\beta^\prime)} > \frac{1}{2}$. Hence
$|\sin{(\beta^\prime-\beta)}| < \frac{1}{2N \frac{1}{2}} =
\frac{1}{N}$. Since $|\frac{\beta^\prime - \beta}{2}| <
|\sin{(\beta^\prime - \beta)}| < |\beta^\prime - \beta|$, we can
write $|\frac{\beta^\prime - \beta}{2}| < \frac{1}{N}$, i.e.,
$|\beta^\prime - \beta| < \frac{2}{N}$.

One can take $\phi(\beta) = \frac{\sin{(2k + 1)\beta}}{\sin{\beta}}$
and use Taylor's series expansion for $\phi(\beta)$ to get the upper
bound on $|\frac{\sin{(2k + 1)\beta^\prime}}{\sin{\beta^\prime}} -
\frac{\sin{(2k + 1)\beta}}{\sin{\beta}}|$. We consider $\phi(\beta +
h) = \frac{\sin{(2k + 1)\beta^\prime}}{\sin{\beta^\prime}}$, where
$h = \beta^\prime - \beta$ is a small quantity. Now $\phi(\beta + h)
- \phi(\beta) \approx h\frac{d\phi(\beta)}{d\beta}$ with error term
bounded by $R_1(\beta)$, the remainder when only the first term in
the Taylor's series is considered. As $k$ increases, the value of
$\beta$ falls in the neighbourhood of $\frac{\pi}{4}$. It can be
checked that $|\frac{d \phi(\beta)}{d \beta}|_{\beta =
\frac{\pi}{4}} < \sqrt{2}(2k+2)$. Also we calculate
$|R_1(\frac{\pi}{4})| = \frac{h^2}{2} |\frac{d^2
\phi(\beta)}{d\beta^2}|_ {\beta = \frac{\pi}{4}+\alpha h} <
\frac{h^2}{2}\sqrt{2((2k+1)^2-1)^2+16}$, where $0 < \alpha < 1$. As
$|\phi(\beta + h) - \phi(\beta)| < |\frac{d \phi(\beta)}{d
\beta}|_{\beta = \frac{\pi}{4}} + |R_1(\frac{\pi}{4})|$, we get
$|\frac{\sin{(2k + 1)\beta^\prime}}{\sin{\beta^\prime}} -
\frac{\sin{(2k + 1)\beta}}{\sin{\beta}}| < \frac{8(k+1)}{N}$.

Since, $\frac{\sin{(2k + 1)\beta}}{\sin{\beta}} = 0$, we have
$\frac{\sin{(2k + 1)\beta^\prime}}{\sin{\beta^\prime}} <
\frac{8(k+1)}{N}$. Thus $a_k(\mu_k^\prime) = \frac{1}{\sqrt{N}}
\frac{\sin{(2k + 1)\beta^\prime}}{\sin{\beta^\prime}} <
\frac{8(k+1)}{N\sqrt{N}}$.

Thus, if $wt(f) = \mu_k N$, then we would have got $x$ such that $f(x) = 1$
with certainty. As $\mu_t N$, may not be an integer, we have considered
functions with $wt(f) = \mu_k^\prime N$ which is an integer such that
$|\mu_k^\prime N - \mu_k N| \leq 0.5$. In this case the probability
of (wrongly) observing an $x$ such that $f(x) = 0$ is
$(a_k(\mu_k^\prime))^2 (1 - \mu_k^\prime) N
< (\frac{8(k+1)}{N\sqrt{N}})^2 (1 - \mu_k^\prime) N
< \frac{64(k+1)^2}{N^2}$.

Similarly, if $wt(f) = (1-\mu_k) N$, then we would have got $x$ such that
$f(x) = 0$ with certainty. As $(1-\mu_k) N$, may not be an integer, we have to
consider functions with $wt(f) = (1-\mu_k^\prime) N$ which is an integer.
In this case the probability of (wrongly) observing an $x$ such that
$f(x) = 1$ is $< \frac{64(k+1)^2}{N^2}$. \qed

Since the function $f$ is available in the form of an oracle, the best
known classical probabilistic algorithm can work as follows. For $k$ many
iterations it can present random inputs to the oracle and guess the
function is of lower weight if the output zero appears more frequently and
guess the function is of higher weight if the output one appears more
frequently. As we consider the majority rule, we choose the number of
iterations as odd in the classical probabilistic algorithm. This will
always guarantee majority of either output zero or output one (in the case
of an even number of iterations there may be the possibility of a tie
and then a random decision has to be taken). For general analysis, the
estimate of probabilities will remain almost the same, and we only present
the analysis when the number of iterations is odd.

Note that for classical lower bound proofs to determine the majority
one may refer to~\cite{LA93}. However, we provide a complete result
with proof which is suitable for our analysis.
Here, the probability of the correct answer in a single step
is $\cos^2{\frac{k}{2k+1}\frac{\pi}{2}}$.
After $g(k)$ many iterations, $g(k)$ in $\Omega(k)$,
probability of success
$p_s = 1 - \sum_{i = 0}^{\frac{g(k)-1}{2}}\comb{g(k)}{i}
(\cos^2{\frac{\pi k}{2(2k+1)}})^i
(\sin^2{\frac{\pi k}{2(2k+1)}})^{g(k)-i}$. Let us now present
the following technical result.
\begin{proposition}
\label{claprop}
For odd positive integer $k$ and $g(k)$ be in $\Omega(k)$, let
$$E(k, g(k)) = \sum_{i = 0}^{\frac{g(k)-1}{2}}\comb{g(k)}{i}
(\cos^2{\frac{\pi k}{2(2k+1)}})^i
(\sin^2{\frac{\pi k}{2(2k+1)}})^{g(k)-i}.$$ Then
\begin{center}
\begin{tabular}{lcl}
$\lim_{k \rightarrow \infty} E(k, k)$ & $=$ & $0.5$,\\
$\lim_{k \rightarrow \infty} E(k, k^2)$ & $>$ & $0.2$ and\\
$\lim_{k \rightarrow \infty} E(k, k^s)$ & $=$ & $0$ for $s > 2$.
\end{tabular}
\end{center}
\end{proposition}
\pf Let $X_k$ follow an identical and independent binomial
distribution having the parameters $g(k), p_k = \cos^2{\frac{\pi
k}{2(2k+1)}}$. Let $\eta_k = E(k, g(k)) = \sum_{i =
0}^{\frac{g(k)-1}{2}}\comb{g(k)}{i} (p_k)^i (1-p_k)^{g(k)-i} =
Prob(X_k \leq \frac{g(k) - 1}{2})$. Consider $X_{k, i} \sim iid \
Bernoulli(p_k)$, where $1 \leq i \leq g(k)$. So, $X_k = X_{k, 1} +
\ldots + X_{k, g(k)}$. Thus, $var(X_k) = \sigma_k^2 = var(X_{k, 1} +
\ldots + X_{k, g(k)}) = g(k) p_k (1 - p_k)$.

Since $g(k) \rightarrow \infty$ as $k \rightarrow \infty$, the Central Limit
Theorem is applicable.
Define $Z_k = \frac{X_k - g(k)p(k)}{\sigma_k}$.
So $\eta_k = Prob(X_k \leq \frac{g(k) - 1}{2}) = Prob(Z_k \leq \zeta_k)$,
where $\zeta_k = \frac{g(k)(\frac{1}{2}-p_k) - \frac{1}{2}}{\sigma_k}$.
Suppose, $\lim_{k \rightarrow \infty} \zeta_k = \zeta \in \mathbb{R}$.
Since convergence in distribution holds in this case,
$\lim_{k \rightarrow \infty} \eta_k = \Phi(\zeta)$, where $\Phi$ is the
cumulative distribution function of the standard normal variate. Hence,
$\zeta = \lim_{k \rightarrow \infty} \zeta_k = \lim_{k \rightarrow \infty}
\frac{g(k)(\frac{1}{2}-p_k) - \frac{1}{2}}{\sigma_k} =
\lim_{k \rightarrow \infty}
\frac{g(k)(\frac{1}{2}-p_k) - \frac{1}{2}}{\sqrt{g(k) p_k (1 - p_k)}}
= \lim_{k \rightarrow \infty} \sqrt{g(k)}(1 - 2p_k)$ as
$\lim_{k \rightarrow \infty} \sqrt{p_k(1-p_k)} = \frac{1}{2}$.
Thus,
\begin{center}
\begin{tabular}{lcl}
$\zeta$ & $=$ & $\lim_{k \rightarrow \infty} \sqrt{g(k)}(-\cos{\frac{\pi k}{2k+1}})$\\
  \     & $=$ & $-\lim_{k \rightarrow \infty} \sqrt{g(k)}\sin{\frac{\pi}{4k+2}}$\\
  \     & $=$ & $-\lim_{k \rightarrow \infty} \frac{\sin{\frac{\pi}{4k+2}}}{\frac{\pi}{4k+2}}\frac{\pi}{4k+2} \sqrt{g(k)}$\\
  \     & $=$ & $ -\lim_{k \rightarrow \infty} \frac{\pi}{4k+2} \sqrt{g(k)}$.
\end{tabular}
\end{center}
So we get, $\zeta = 0$ when $g(k) = t$,
$\zeta = -\frac{\pi}{4}$ when $g(k) = k^2$ and
$\zeta = -\infty$, when $g(k) = k^s, s > 2$.
Hence,
$\lim_{k \rightarrow \infty} \eta_k$ is
$= \Phi(0) = 0.5$, when $g(k) = k$,
$= \Phi(-\frac{\pi}{4}) > 0.2$ when $g(k) = k^2$ and
$= \Phi(-\infty) = 0$ when $g(k) = k^s, s > 2$.
This gives the proof. \qed

\begin{theorem}
\label{clathm} Consider Boolean functions on $n$ variables and let
$N = 2^n$. After $k$ iterations, the best known classical
probabilistic algorithm can distinguish two Boolean functions of
weights $\lfloor{N\sin^2{\frac{k}{2k+1}\frac{\pi}{2}}}\rceil$ and
$\lfloor{N\cos^2{\frac{k}{2k+1}\frac{\pi}{2}}}\rceil$ with success
probability $p_s = 0.5$ when $g(k) = k$, $p_s < 0.8$ when $g(k) =
k^2$ and $p_s = 1$ when $g(k) = k^s, s > 2$, for large $k$.
\end{theorem}
\pf The proof follows from
$p_s = 1 - \sum_{i = 0}^{\frac{g(k)-1}{2}}\comb{g(k)}{i}
(\cos^2{\frac{\pi k}{2(2k+1)}})^i
(\sin^2{\frac{\pi k}{2(2k+1)}})^{g(k)-i}$ and the results in
Proposition~\ref{claprop}. \qed

Based on the results of Theorem~\ref{mainthm} and Theorem~\ref{clathm}, it is
clear that when the quantum algorithm can achieve a success probability almost
unity, then the best known classical algorithm can achieve a success probability
almost $0.5$ (from above) after $k$ many steps. The classical
algorithm can achieve a success probability almost unity only after $k^s$
many steps for $s > 2$. Thus the quantum algorithm can achieve a quadratic
speed up in this case.

\subsection{Distinguishing Boolean functions from two different Sets of Weights}\label{difwt}

\begin{table*}
\begin{center}
{\tiny
\begin{tabular}{|l|l|}
\hline
$k$ & The roots of $a_k = 0$ (top line) and $b_k = 0$ (bottom line)\\
\hline
1 & 0.250000, \\
 \ & 0.750000, \\
 \hline
 2 & 0.095492, 0.654508, \\
 \ & 0.345492, 0.904508, \\
 \hline
 3 & 0.049516, 0.388740, 0.811745, \\
 \ & 0.188255, 0.611260, 0.950484, \\
 \hline
 4 & 0.030154, 0.250000, 0.586824, 0.883022, \\
 \ & 0.116978, 0.413176, 0.750000, 0.969846, \\
 \hline
 5 & 0.020254, 0.172570, 0.428843, 0.707708, 0.920627, \\
 \ & 0.079373, 0.292292, 0.571157, 0.827430, 0.979746, \\
 \hline
 6 & 0.014529, 0.125745, 0.322698, 0.560268, 0.784032, 0.942728, \\
 \ & 0.057272, 0.215968, 0.439732, 0.677302, 0.874255, 0.985471, \\
 \hline
 7 & 0.010926, 0.095492, 0.250000, 0.447736, 0.654508, 0.834565, 0.956773, \\
 \ & 0.043227, 0.165435, 0.345492, 0.552264, 0.750000, 0.904508, 0.989074, \\
 \hline
 8 & 0.008513, 0.074891, 0.198683, 0.363169, 0.546134, 0.722869, 0.869504, 0.966236, \\
 \ & 0.033764, 0.130496, 0.277131, 0.453866, 0.636831, 0.801317, 0.925109, 0.991487, \\
 \hline
 9 & 0.006819, 0.060263, 0.161359, 0.299152, 0.458710, 0.622743, 0.773474, 0.894570, 0.972909, \\
 \ & 0.027091, 0.105430, 0.226526, 0.377257, 0.541290, 0.700848, 0.838641, 0.939737, 0.993181, \\
 \hline
10 & 0.005585, 0.049516, 0.133474, 0.250000, 0.388740, 0.537365, 0.682671, 0.811745, 0.913119, 0.977786, \\
 \ & 0.022214, 0.086881, 0.188255, 0.317329, 0.462635, 0.611260, 0.750000, 0.866526, 0.950484, 0.994415, \\
 \hline
\end{tabular}
}
\end{center}
\caption{Roots of $a_k = 0$ and $b_k = 0$.}
\label{tab1}
\end{table*}

Let us consider two different sets $A_k$ and $B_k$, where $A_k$
contains the roots of $a_k = 0$ and $B_k$ contains the roots of $b_k
= 0$. From Proposition~\ref{prop1}, it is clear that $A_k =
\{\sin^{2}{\frac{(2m - 1)}{(2k+1)}\frac{\pi}{2}} | m = 1, \ldots,
k\}$ and $B_k = \{\cos^{2}{\frac{(2m - 1)}{(2k+1)}\frac{\pi}{2}} | m
= 1, \ldots, k\}$. One can see Table~\ref{tab1} for a few examples.

Given any $k$,
Algorithm~\ref{algo:randomized_weight_decision_algorithm} can
distinguish whether a Boolean function is either from the weights
$\lfloor N\sin^{2}{\frac{(2m - 1)}{(2k+1)}\frac{\pi}{2}} \rceil$ or
from the weights $\lfloor N\cos^{2}{\frac{(2m -
1)}{(2k+1)}\frac{\pi}{2}} \rceil$, with good success probability for
$m = 1, \ldots, k$. This also helps in solving the question can we
distinguish two Boolean functions of weights $Nw_1$ and $Nw_2$ where
$w_1 + w_2 \neq 1$? As example, one can check the case for $k = 10$
in Table~\ref{tab1}, where we may be able to distinguish Boolean
functions with weights $Nw_1$ and $Nw_2$, for $w_1 \approx 0.388740$ and
$w_2 \approx 0.317329$. However, we leave this for future research as in
this paper we mainly focus on distinguishing functions having
weights $Nw$ and $N(1-w)$.

\section{Sure Success Weight Decision Algorithm}

\subsection{Motivation}

By Theorem \ref{mainthm}, we know that the Algorithm
\ref{algo:randomized_weight_decision_algorithm} is a randomized one 
and not sure success algorithm. In the Grover-like database search algorithms, 
number of ideas have been proposed to achieve sure success. We try to
exploit similar strategies here, though we need to make certain subtle
modifications for this purpose.

\subsection{Brassard's Sure Success Database Search}
\begin{figure}[t]
\centering%
\subfigure[Hilbert Space] { \label{fig:grover-hilbert-space}
\includegraphics[scale=0.85]{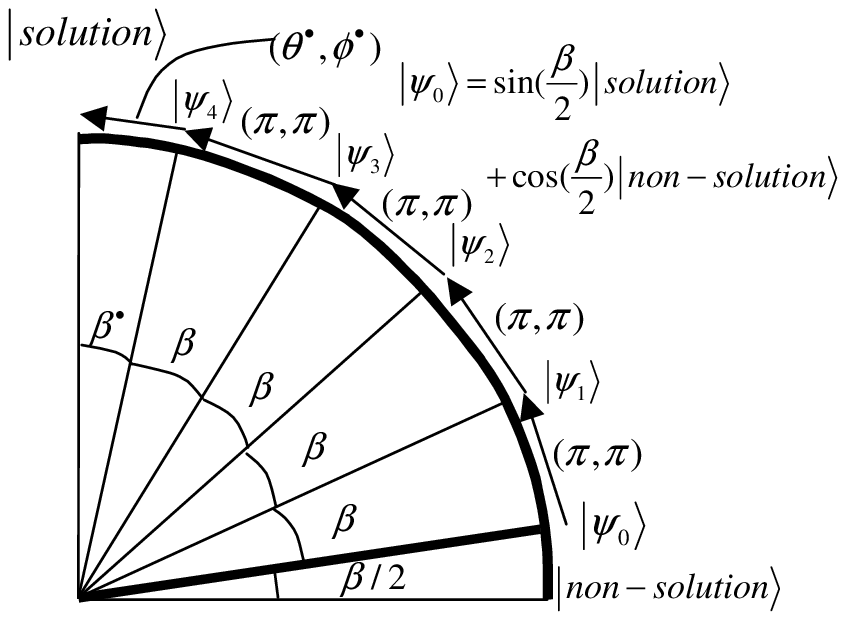}
}%
\quad   \quad%
\subfigure[Bloch Sphere] {
\label{fig:grover-bloch-sphere}%
\includegraphics[scale=0.55]{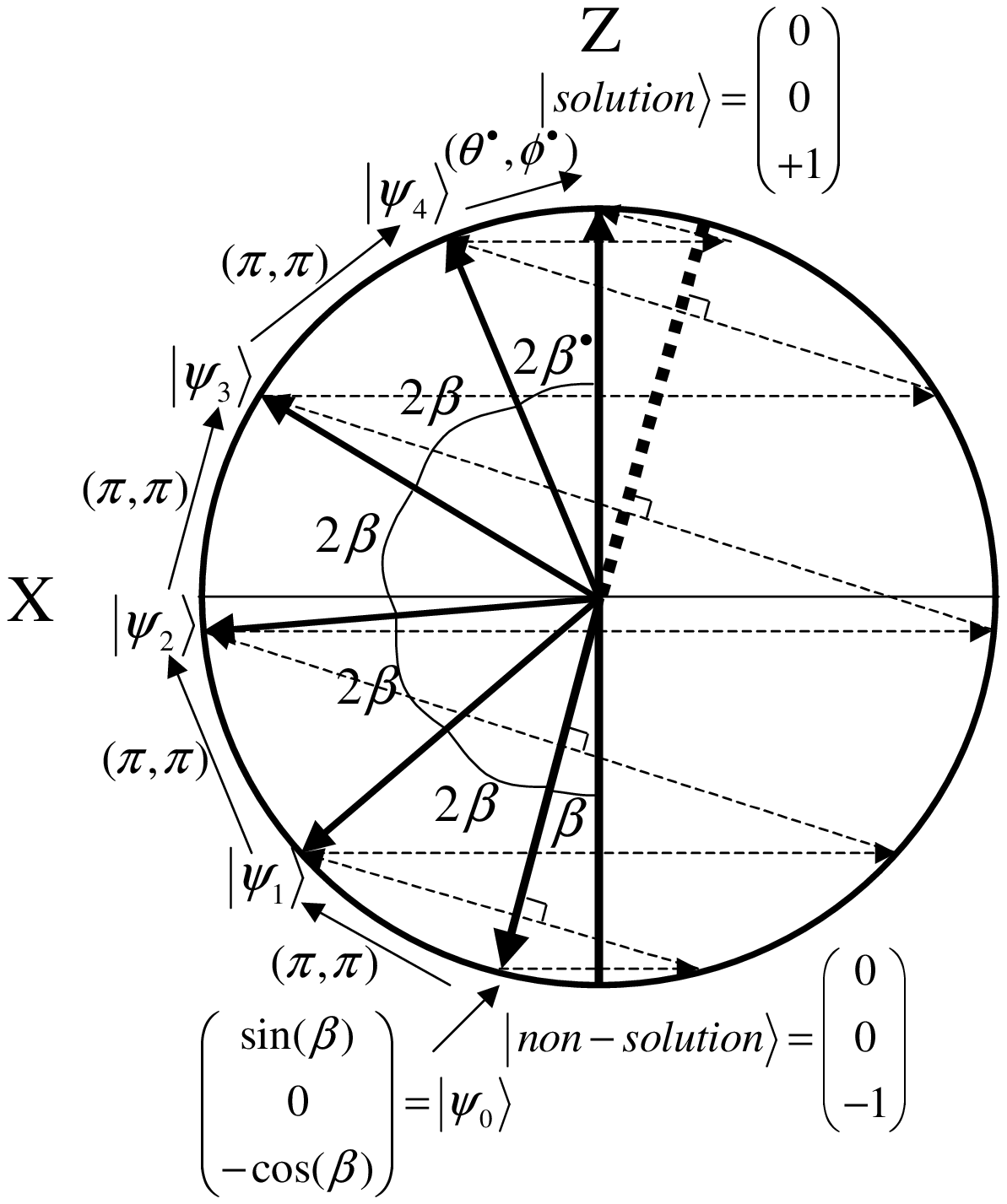}
}%
\caption{Brassard's Sure Success Database Search}
\end{figure}

Many algorithms have been proposed for a sure success database
search algorithm based on Grover search
\cite{HSIEH-LI,LI-HWANG-HSIEH-WANG,LONG-LI-SUN,HOYER,LONG,2000quant.ph..5055B}.
Meanwhile, we need to reformulate and to generalize the original
Grover operators, $G_n$ and $U_f$, for a sure success method as
follows.
\begin{center}
$G_n = -I_{|\psi_0\rangle}(\theta) = -\{I-(1-e^{i\theta})|\psi_0\rangle\langle\psi_0|\}$.\\
$U_f = I_{|solution\rangle}(\phi) = I-(1-e^{i\phi})|solution\rangle\langle solution|$.\\
\end{center}
A sure success database search can find one of solutions
exactly by changing two phases, $\theta$ and $\phi$ of
$I_{|\psi_0\rangle}(\theta)$ and $I_{|solution\rangle}(\phi)$. Can
we apply this kind sure success approach in the Grover database
search into our problem? We try to exploit 
Brassard method \cite{2000quant.ph..5055B} for this. The method is based on
the following approach. The required minimum number of Grover
operation is calculated, which assumes to be $k$. Then, from
$1^{st}$ to ${(k-1)}^{th}$ operation, the original Grover operation
is applied. However, for the last $k^{th}$ operation, a slightly
different Grover operation is applied by controlling two phases,
$\theta$ and $\phi$ for $I_{|\psi_0\rangle}(\theta)$ and
$I_{|solution\rangle}(\phi)$. Figures \ref{fig:grover-hilbert-space}
and \ref{fig:grover-bloch-sphere} show the Brassard's method in the
Hilbert space and Bloch sphere, respectively. Please note that the
inversion operators,
$-I_{|\psi_0\rangle}(\theta)I_{|solution\rangle}(\phi)$ in the
Hilbert space, correspond to the rotation operators,
$-e^{i(\frac{\theta}{2}+\frac{\phi}{2})}R_{|\psi_0\rangle}(-\theta)R_{|solution\rangle}(-\phi)$
in the Bloch sphere \cite{LONG-TU-LI-ZHANG-YAN}. As shown in the
figures, the last step of the Grover operation uses different phases
for two inversions in the Hilbert space and for two rotations in the
Bloch Sphere.

\subsection{Approach}
At the first sight, it looks that Brassard method can be directly applied
for the weight decision problem. However, it is a little bit different
from the Brassard's case. Brassard method changes
only the last step because the goal of this approach is to rotate
the $(k-1)^{th}$ state to the solution state. However, in the weight
decision case, we need to satisfy that two initial states for
different weights should be rotated to the solution and the
non-solution state exclusively after operations. In other words, if
the proposed method rotates the initial state for $wN$ weight case
to the solution state, the same operation should rotate the initial
state for $(1-w)N$ weight case to the non-solution state. By this
condition, we can make a relation between the initial states for
$wN$ and $(1-w)N$ to the solution and non-solution states. Finally,
we can decide the weight exactly from measuring the final state and
evaluating the function with the measured value $\hat x$. To satisfy
this condition, we proposed a method which changes the very last two
operations. As a result, from $1^{st}$ to ${(k-2)}^{th}$ operation,
we use the ($\pi$,$\pi$) phase angles for the
$I_{|\psi_0\rangle}(\theta)$ and $I_{|solution\rangle}(\phi)$
operations. However, for ${(k-1)}^{th}$ and $k^{th}$ operation,
($-\theta_1$, $\pi$) and ($-\theta_2$, $\pi$) phases should be used,
respectively. Algorithm
\ref{algo:sure_success_weight_decision_algorithm} describes the
overall idea.

\begin{algorithm} {\em Sure Success Weight Decision Algorithm}
\label{algo:sure_success_weight_decision_algorithm}
\begin{center}
\begin{tabular}{|ll|} \hline
1.  & {\rm $|\psi_0\rangle = |0\rangle^{\otimes n} |1\rangle$, $i=0$,}\\
    & {\rm if $0 < Nmin(w,1-w) \leq Nsin^2\frac{\pi}{5}$, $k$ is 2,}\\
    & {\rm otherwise, $k$ satisfies $Nsin^2{\frac{k-1}{2k-1}\frac{\pi}{2}} < Nmin(w,1-w) \leq Nsin^2{\frac{k}{2k+1}\frac{\pi}{2}}$}.\\
    & $\frac{k-1}{2k-1}\pi< \beta_{for\,smaller\,weight} \leq \frac{k}{2k+1}\pi$. \\
    & $\pi-\frac{k-1}{2k-1}\pi > \beta_{for\,bigger\,weight} \geq \pi-\frac{k}{2k+1}\pi$. \\
2.  & {\rm while($i<(k-2)$) do} \\
    & \{\\
    & \hspace{20pt} $|\psi_{i+1}\rangle=-I_{|\psi_0\rangle}(\pi)I_{|solution\rangle}(\pi)|\psi_{i}\rangle$\\
    & \hspace{20pt} $i = i+1$\\
    & \}\\
3.  & $|\psi_{k-1}\rangle=-I_{|\psi_0\rangle}(-\theta_1)I_{|solution\rangle}(\pi)|\psi_{k-2}\rangle$\\
4.  & $|\psi_{k}\rangle=-I_{|\psi_0\rangle}(-\theta_2)I_{|solution\rangle}(\pi)|\psi_{k-1}\rangle$\\
5.  & {\rm measure $|\psi_{k}\rangle$ in the computational basis.} \\
    & {\rm let the result be ${\hat x}$.}\\
6-1.& {\rm if $k$ is odd and if $f({\hat x})$ = 0 then} \\
    & {\rm \hspace{20pt} $wt(f)=Nmax(w,1-w)$}\\
    & {\rm else $wt(f)=Nmin(w,1-w)$.}\\
6-2.& {\rm if $k$ is even and if $f({\hat x})$ = 1 then} \\
    & {\rm \hspace{20pt} $wt(f)=Nmax(w,1-w)$}\\
    & {\rm else $wt(f)=Nmin(w,1-w)$.}\\
\hline
\end{tabular}
\end{center}
\end{algorithm}

\subsection{Alternative Final State}
In the Brassard database search algorithm, the final state should be
the solution state. However, in the weight distinction case, two
final states after $k$ operations should be located in the solution
and non-solution state exclusively. Meanwhile, there are no
restrictions of the locations of final states except that two final
states should be located at solution and non-solution states
exclusively. Hence the locations of two final states can be
alternatively changed with the required number of operations. As a
result, we need to think about the final states of the proposed
method. The previously proposed method, Algorithm
\ref{algo:randomized_weight_decision_algorithm}, is a special case
where the ($\pi$, $\pi$) phase angles are used for the
$I_{|\psi_0\rangle}(\theta)$ and $I_{|solution\rangle}(\phi)$
operations. Hence we can infer that the final state for the smaller
weight case after $k$ operations in the special weight decision
algorithm, in Bloch sphere, is
\begin{equation}\label{final_state}
\left( {{\begin{array}{*{20}c}
 {\sin (2k + 1)\beta } \hfill \\
 0 \hfill \\
 { - \cos (2k + 1)\beta } \hfill \\
\end{array} }} \right) ,
\end{equation}
where $\beta = \frac{k}{2k + 1}\pi$. Note that all state vectors are
represented in the Bloch sphere hereafter. Therefore, we can know
that the final state for the smaller wight is alternatively changed
with the required $k$ operations such as
\begin{equation}\label{alternative_final_state}
\left( {{\begin{array}{*{20}c}
 {\sin ((2k + 1)\frac{k}{2k + 1}\pi )} \hfill \\
 0 \hfill \\
 { - \cos ((2k + 1)\frac{k}{2k + 1}\pi )} \hfill \\
\end{array} }} \right) = \left( {{\begin{array}{*{20}c}
 {   \sin (k\pi )} \hfill \\
 0 \hfill \\
 { - \cos (k\pi )} \hfill \\
\end{array} }} \right) = \left\{ {\begin{array}{l}
 \left( {{\begin{array}{*{20}c}
 0 \hfill \\
 0 \hfill \\
 { + 1} \hfill \\
\end{array} }} \right) $for odd$ \,k \\
 \left( {{\begin{array}{*{20}c}
 0 \hfill \\
 0 \hfill \\
 { - 1} \hfill \\
\end{array} }} \right) $for even$ \,k  \\
 \end{array}} \right. .
\end{equation}

Meanwhile, the $(0,0,+1)^t$ and $(0,0,-1)^t$ states in Bloch sphere
represent solution and non-solution states in Hilbert space,
respectively. In summary, if the required number of $k$ is odd, the
final state for the smaller (bigger) weight case should be located in
the solution (non-solution) state. On the other hand, if the required number
of $k$ is even, the final state for the smaller (bigger) weight case
should be located in the non-solution (solution) state. On the other
hand, for the bigger weight case, we can easily analyze the
alternative final state based on the number of operations with the
initial angle, $\pi - \beta$. This analysis means that we need to
find two phase conditions with the required number of operations.

\subsection{Modification of Last Two Operations}
\begin{figure}[t]
\centering
\subfigure{\includegraphics[scale=0.80]{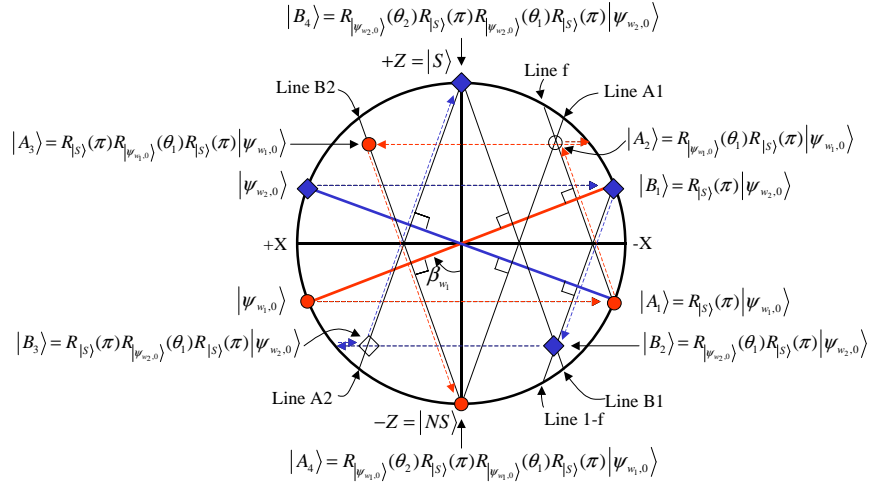}}
\caption{Last Two Operations} \label{fig:last-two-operations}
\end{figure}

Figure \ref{fig:last-two-operations} explains how we can rotate two
initial states to different final states, which should be located in
the different poles in Bloch sphere as solution,$(0,0,+1)^t$, and
non-solution,$(0,0,-1)^t$, state. Note that Figure
\ref{fig:last-two-operations} shows the case when only two
operations are sufficient to decide the exact weight. In the figure,
the circle and diamond mean two states, which have the smaller and
the bigger weight, respectively. Our purpose is to find phase
conditions, which can rotate two initial states to different poles
exclusively with the same phase conditions. Hence if the weight is the
smaller (bigger) one and the method rotates the initial state to the
solution state, the method should rotate the initial state of the
bigger (smaller) weight case to the non-solution state. Note that all
figures hereafter are viewed from $+Y$ direction in the Bloch sphere
for easy understanding. Therefore, only $X$ and $Z$ directions and
the $X-Z$ plane are shown in all the figures. In the initial step, two
initial states $|A_0\rangle$ and $|B_0\rangle$ are
$(\sin\beta,0,-\cos\beta)^t$ and
$(\sin(\pi-\beta),0,-\cos(\pi-\beta))^t=(\sin\beta,0,\cos\beta)^t$,
respectively. At the first step, two initial states are rotated to
$|A_1\rangle$ and $|B_1\rangle$ states by using rotation
$R_{|solution\rangle }(\pi)$. Therefore, only the sign of $x$ is
changed. In the second step, two last states are rotated to
$|A_2\rangle$ and $|B_2\rangle$ states by using rotation
$R_{|\psi_0\rangle }(\theta_1)$. Meanwhile, $|A_1\rangle$ state
should be rotated to $|A_2\rangle$ state, which is the cross point
between line $A1$ and line $f$. The line $f$ is a path, where the
point $|A_1\rangle$ can be rotated by the rotation
$R_{|\psi_0\rangle }(\theta_1)$. As the same rule, the same rotation
operator rotates $|B_1\rangle$ state to $|B_2\rangle$ state, where
$|B_2\rangle$ state is the cross point between line $B1$ and line
$1-f$. Because two last states are rotated with the same phase
angle, the value of $x$ is the same, but values of $y$ and $z$ have
different sign values between them. At the third step,
$R_{|solution\rangle }(\pi)$ operator is used for the last two
states. Therefore, the signs of $x$ and $y$ are changed as shown in
$|A_3\rangle$ and $|B_3\rangle$ states. At the last step, the two
final states are rotated to two different poles exclusively by using
the same rotation angle, $\theta_2$. In other words, $|A_3\rangle$
state moves to the $-Z$ pole, non-solution state, and $|B_3\rangle$
to the $+Z$ pole, solution state. Finally, if we measure the final
state and the measured value, then $\hat x$, is one of
solutions (non-solutions) and we can decide that the weight is
bigger (smaller) one. Meanwhile, the key point of this approach is to
find two cross points, $|A_2\rangle$ and $|B_2\rangle$ with the
required number of operations.

\subsection{Correctness}
In the proposed method, we have to change two phases only for the
last two operations, not for other operations because until
$(k-2)^{th}$ operation, there is no cross point such as
$|A_2\rangle$ and $|B_2\rangle$ in Figure
\ref{fig:last-two-operations}. Therefore, we need to show that until
$(k-2)^{th}$ operation, there are no cross point, but in the
$(k-1)^{th}$ operation, there are two cross points. For easy
explanation, we only consider when the required number of operation
is odd case and the weight is smaller one. Hence $k$ is odd and
$\frac{k-1}{2k-1}\pi< \beta \leq \frac{k}{2k+1}\pi$. Other cases can
be proved with the same approach.

\subsubsection{No Cross Point until $(k-2)^{th}$ Operation}
Figure \ref{fig:until-k-2-operations} shows the last state,
$|\psi_{k-2}\rangle$, when $(k-2)$ operations of $R_{|\psi_0\rangle
}(\pi)R_{|solution\rangle }(\pi)$ are applied. Note that in the odd
$k$ case, $|\psi_{k-2}\rangle$ state should be located in the right
upper part, i.e., $(-X,+Z)$ area. Meanwhile, the line $A1$, which is the
line perpendicular to the axis of $|\psi_{1-w}\rangle$ and meets the
south pole, is $Z = -X\tan\beta-1$. The value of $x$ of the cross
point between the line $A1$ and the circle is $x_{line} =
-\sin2\beta$, and the value of $x$ of $|\psi_{k-2}\rangle$,
$x_{k-2}$, is $\sin(2(k-2)+1)\beta$. Therefore, to show that there
is no cross point until $(k-2)^{th}$ operation, we need to prove
that $x_{line}$ is always larger than $x_{k-2}$.

\begin{theorem}[No Cross Point until $(k-2)^{th}$ Operation] \label{no_cross_point}
$- sin2\beta > sin(2(k-2)+1)\beta$, where $\frac{k-1}{2k-1}\pi<
\beta \leq \frac{k}{2k+1}\pi$.
\end{theorem}
\begin{proof}
From the value of $\beta$, we can get the value of $-\sin2\beta$ as
$-\sin(\frac{\pi}{2k-1}) < -\sin2\beta \leq
-\sin(\frac{\pi}{2k+1})$. Meanwhile, $(k-2)\pi+\frac{\pi}{2k-1} <
(2k-3)\beta \leq (k-2)\pi + \frac{2\pi}{2k+1}$. Because $k$ is odd
in this case, $\sin((k-2)\pi+\frac{\pi}{2k-1})>\sin(2k-3)\beta \geq
\sin((k-2)\pi+\frac{2\pi}{2k+1})$. Finally, $-\sin(\frac{\pi}{2k-1})
> \sin(2k-3)\beta \geq -\sin(\frac{2\pi}{2k+1})$. Therefore,
$-\sin2\beta
> \sin(2k-3)\beta$.
\end{proof}

\subsubsection{First Cross Point in $(k-1)^{th}$ Operation}

\begin{figure}[t]
\centering%
\subfigure[No Cross until $(k-2)^{th}$ Operation] {
\label{fig:until-k-2-operations}
\includegraphics[scale=0.70]{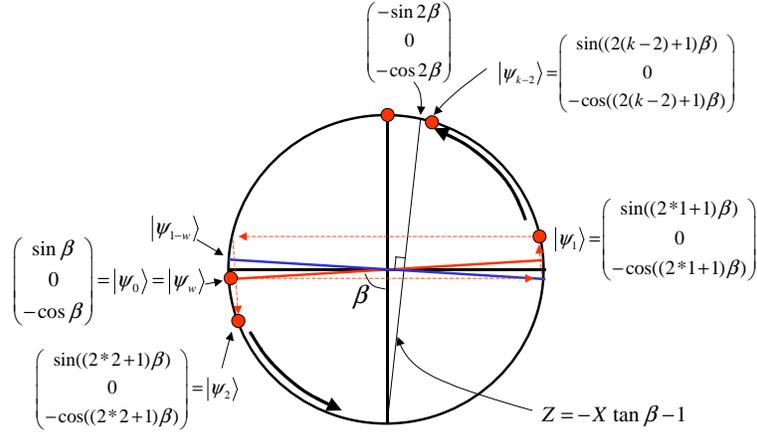}
}%
\quad   \quad%
\subfigure[First Cross in $(k-1)^{th}$ Operation] {
\label{fig:first-k-1-operation}%
\includegraphics[scale=0.70]{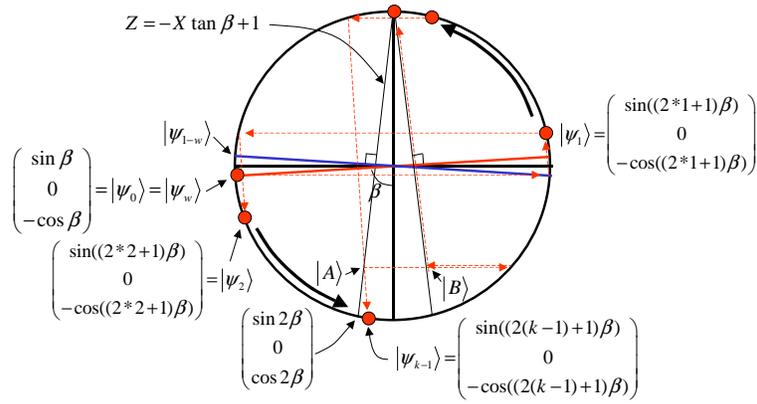}
}%
\caption{Correctness of Modification of Last Two Operations}
\end{figure}

Figure \ref{fig:first-k-1-operation} explains why, in the
$(k-1)^{th}$ operation, there is the first cross point between the
line $A1$ and the line $f$. To prove this, we need to show that the
value of $x$ of the cross point, $x_{line}$, between the line $A1$
and the circle is always larger than equal to the value of $x$ of
$|\psi_{k-1}\rangle$ state, $x_{k-1}$. Note that the value of
$x_{line}$ is $\sin2\beta$, and the value of the $x_{k-1}$ is
$\sin(2(k-1)+1)\beta$.

\begin{theorem}[First Cross Point in $(k-1)^{th}$ Operation] \label{first_cross_point}
$\sin2\beta \geq \sin(2(k-1)+1)\beta$, where $\frac{k-1}{2k-1}\pi<
\beta \leq \frac{k}{2k+1}\pi$.
\end{theorem}
\begin{proof}
From Theorem \ref{no_cross_point}, we can get the value of
$\sin2\beta$ as $\sin(\frac{\pi}{2k-1})>\sin2\beta\geq
\sin(\frac{\pi}{2k+1})$. Meanwhile, $(k-1)\pi<(2k-1)\beta\leq
(k-1)\pi + \frac{\pi}{2k+1}$. Because $k$ is odd in this case,
$\sin(k-1)\pi < \sin(2k-1)\beta \leq
\sin((k-1)\pi+\frac{\pi}{2k+1})$. Finally, $0<\sin(2k-1)\beta\leq
\sin(\frac{\pi}{2k+1})$. Therefore, $\sin2\beta \geq
\sin(2k-1)\beta$.
\end{proof}

\subsection{Phase Conditions}
\begin{figure}[t]
\centering \subfigure{\includegraphics[scale=0.70]{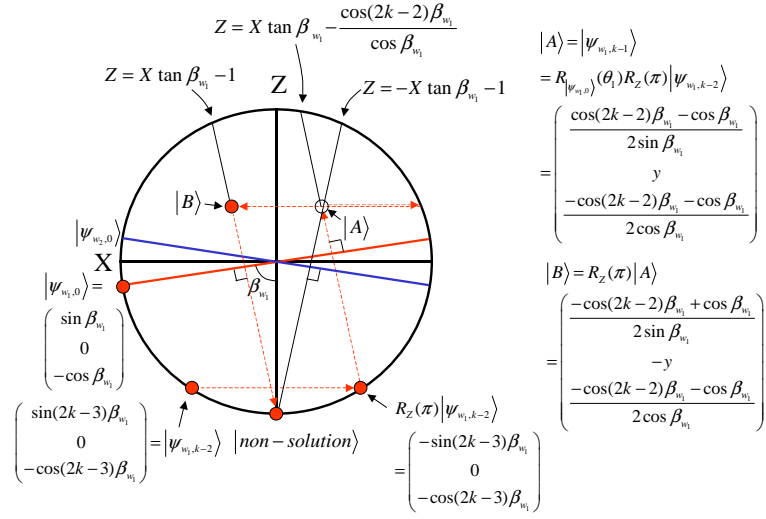}}
\caption{Last Two Operations for Even $k$} \label{fig:even-weight}
\end{figure}

\begin{figure}[t]
\centering \subfigure{\includegraphics[scale=0.70]{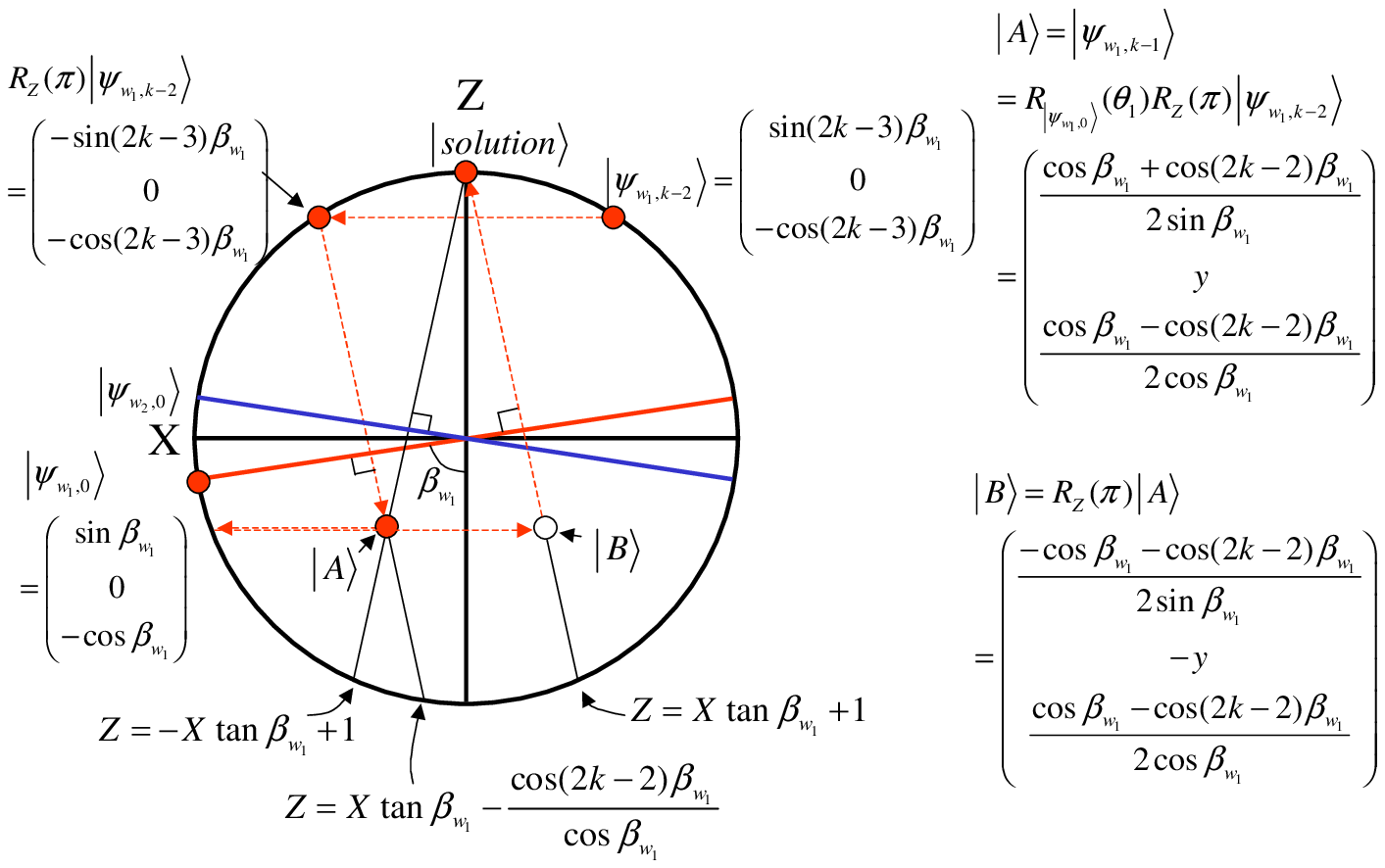}}
\caption{Last Two Operations for Odd $k$} \label{fig:odd-weight}
\end{figure}

Figure \ref{fig:even-weight} and \ref{fig:odd-weight} show the trace
of states for the last two operations when $k$ is even and odd,
respectively. Note that we only consider the Boolean function with
the smaller weight because the phase conditions are the same for the
bigger weight case. At first, to rotate $|\psi_{k-2}\rangle$ state
to the first cross point $|A\rangle$, $\theta_1$ should satisfy the
following equation
\begin{equation}\label{odd-theta_1_condition}
R_{\left| \psi \right\rangle } (\theta _1 )\left(
{{\begin{array}{*{20}c}
 { - \sin (2k - 3)\beta } \hfill \\
 0 \hfill \\
 { - \cos (2k - 3)\beta } \hfill \\
\end{array} }} \right) = \left( {{\begin{array}{*{20}c}
 {\frac{\cos (2k - 2)\beta - ( - 1)^k\cos \beta }{2\sin \beta }} \hfill \\
 y \hfill \\
 {\frac{ - \cos (2k - 2)\beta - ( - 1)^k\cos \beta }{2\cos \beta }} \hfill
\\
\end{array} }} \right)
.
\end{equation}

As a result, $\theta_1$ should be chosen as a value to satisfy the
following equation
\begin{equation}\label{odd_theta_1}
\cos \theta _1 = \frac{( - 1)^k\cos \beta - \cos 2\beta \cos (2k -
2)\beta }{\sin 2\beta \sin (2k - 2)\beta }.
\end{equation}

Meanwhile, the value of $y$ for the state $|A\rangle$ is calculated
by $y = \sin \theta _1 \sin (2k - 2)\beta$. With the very similar
approach we can find a condition for the value of $\theta_2$, which
should satisfy the following equation
\begin{equation}\label{odd-theta_2_condition}
R_{\left| \psi \right\rangle } (\theta _2 )\left(
{{\begin{array}{*{20}c}
 {\frac{ - \cos (2k - 2)\beta + ( - 1)^k\cos \beta }{2\sin \beta }} \hfill
\\
 { - y} \hfill \\
 {\frac{ - \cos (2k - 2)\beta - ( - 1)^k\cos \beta }{2\cos \beta }} \hfill
\\
\end{array} }} \right) = \left( {{\begin{array}{*{20}c}
 0 \hfill \\
 0 \hfill \\
 { - ( - 1)^k} \hfill \\
\end{array} }} \right)
.
\end{equation}

Finally, $\theta_2$ can be chosen, which satisfies the following
equation
\begin{equation}\label{odd_theta_2}
\cos \theta _2 = \frac{( - 1)^k\sin 2\beta (y\sin \theta _2 - ( -
1)^k\sin \beta )}{\cos \beta \cos 2\beta - ( - 1)^k\cos (2k -
2)\beta } .
\end{equation}

\section{Comparison with Quantum Counting Method}
To argue the efficiency of our algorithm, let us refer to the existing works on
quantum counting \cite{BO98,1998quant.ph..5082B,2000quant.ph..5055B}. The
existing algorithm exploits the period information of Grover iterations. From
this period information, one can guess the number of solutions.
Hence, as like Shor's factoring algorithm, this is the task to find
the period of Grover operations using quantum Fourier transform.

Because we proposed a way to decide the weight from given special
two weights, it is meaningful to compare time complexity between our
approach and the method based on quantum counting with the same
promise.

\subsection{Quantum Counting}
Counting problem is to find the number of solutions of a given
Boolean function. Meanwhile, Grover operation shows some kind period
patterns with the iteration numbers. By this analysis, we can count
the number of solutions with quantum Fourier transform as shown in
Algorithm \ref{algo:quantum_counting}.

\begin{algorithm} {\em Quantum Counting \cite{BO98,1998quant.ph..5082B,2000quant.ph..5055B}}
\label{algo:quantum_counting}
\begin{center}
\begin{tabular}{|ll|} \hline
Let &   ${\mathbf{C}}_F : |m\rangle \otimes |\Psi\rangle \mapsto
|m\rangle \otimes ({\mathbf G}_F)^m
|\Psi\rangle$ \\
Let &   ${\mathbf F}_P : |k\rangle \mapsto \frac{1}{\sqrt{P\;\!}}
              \sum^{P-1}_{l=0} e^{2 \pi \imath kl/P} \,|l\rangle$ \\
1.  &   $|\Psi_0\rangle  \ \leftarrow   {\mathbf W} \otimes
{\mathbf W}|0\rangle|0\rangle $\\
2.  &   $|\Psi_1\rangle  \ \leftarrow  {\mathbf C}_F
\,|\Psi_0\rangle $\\
3.  &   $|\Psi_2\rangle  \ \leftarrow |\Psi_1\rangle$ after the
second register is measured ({\em optional}\,) \label{stp:measure}\\
4.  &   $|\Psi_3\rangle  \ \leftarrow  {\mathbf F}_P \otimes
{\mathbf I} |\Psi_2\rangle $ \label{stp:fourier}\\
5.  &   $\tilde{f}  \ \leftarrow  $  measure $|\Psi_3\rangle$
\quad (if $\tilde{f} >P/2$ then $\tilde{f}\leftarrow (P-\tilde{f})$)\\
6.  &   output: $\ N \sin^2(\tilde{f} \pi/P)$ (and $\tilde f$ if
needed)\\
\hline
\end{tabular}
\end{center}
\end{algorithm}

\subsection{Exploitation and Analysis of Quantum Counting for Weight
of Boolean Function}
We analyze the quantum counting algorithm for three purposes. First,
when a weight $w$ is given, how we can check the correctness of the
given weight with how much time complexity. Second, when two weights
are given, how we can decide the real weight with how much time
complexity. Third, when $n$ possible weights are given, how we can
find the real weight with how much time complexity.

\subsubsection{Check Correctness of a Given Weight $w$}
If a weight is given as $w$=sin$^2\theta$, how can we check whether
this is correct or incorrect and what about time complexity? Let $f=
P \theta / \pi$. If $f$ were an integer, there would be two
possibilities: either $f=0$ (which happens if $t=0$ or $t=N$), in
which case $|\Psi_3\rangle = |0\rangle$, or $t>0$, in which case
$|\Psi_3\rangle = a|f\rangle + b|P-f\rangle$, where $a$ and $b$ are
complex numbers of norm $1/\sqrt{2}$
\cite{BO98,1998quant.ph..5082B,2000quant.ph..5055B}. In other words,
if we assume the value of $P$ as $k\pi/\theta$, the measured value
of $\tilde{f}$ should be $k$. As a result, we can easily check
whether the given weight is correct or not by measuring
$|\Psi_3\rangle$. If the measured value is $k$, the given weight is
correct, otherwise incorrect. In this case, the time complexity is
$O(P)$. Meanwhile, because we already know the value of $\theta$ in
the initial time, we can find the smallest integer value of $P$ as
$k\frac{\pi}{\theta}$. If $\theta$ is $\frac{\pi}{a}$, then $P$ is
just $ka$. Therefore, the time complexity of this case is $O(ka)$
when $\theta$ is $\frac{\pi}{a}$.

\subsubsection{Decide Real Weight $w$ from Two Given Weights $w_1$
and $w_2$}
From the previous section, we can know that when
$w_1=sin^2\theta_1$($w_2=sin^2\theta_2$) is given, we can find the
required number $P$ for $w_1$($w_2$) and the expected measured value
as $k_1$($k_2$). However, when we want to decide which weight is
real one, $k_1$ and $k_2$ should be different because they are the
clues to distinguish. Therefore, we need to find the integer value
$P$, which will be used for both two cases. Two values of $P$ are
$P_1 = k_1\pi/\theta_1$ and $P_2=k_2\pi/\theta_2$. Because we need
to execute the algorithm only once, $P_1$ and $P_2$ should be the
same. Hence, $k_1\pi/\theta_1$ should be equal to $k_2\pi/\theta_2$
and our job is to find suitable $k_1$ and $k_2$. Meanwhile, if
$\theta_1 = \pi/a_1$ and $\theta_2 = \pi/a_2$, $P$ should be $k_1a_1
= k_2a_2$. Considering the time complexity, we need to find the
smallest integer value of $P$ by $P_s=LCM(a_1,a_2)$, where $LCM$ is
the least common integer multiplier. Then, $k_1 = P_s/a_1$ and $k_2
= P_s/a_2$. Finally, with the value of $P_s$, we evaluate the
algorithm, and if the measured value of $\tilde{f}$ is $k_1$, we can
say that the weight is $w_1$. If the measured value of $\tilde{f}$
is $k_2$, we can say that the weight is $w_2$. Time complexity of
this case is $LCM(a_1,a_2)$, where $\theta_1$ is $\pi/a_1$ and
$\theta_2$ is $\pi/a_2$.

Now we need to compare our method and the above method based on
quantum counting when $w_1+w_2 = 1$. If $w_1 =
sin^2(\frac{\pi}{\frac{4k+2}{k}})$ and $w_2 =
sin^2(\frac{\pi}{\frac{4k+2}{k+1}})=cos^2(\frac{\pi}{\frac{4k+2}{k}})$,
our method, Algorithm
\ref{algo:randomized_weight_decision_algorithm} and
\ref{algo:sure_success_weight_decision_algorithm} can decide which
one is real one in $k$ many steps. On the other hand, the
method based on quantum counting needs $P$ as $4k+2$ and if
$\tilde{f}$ is $k$, the real weight is $w_1$ and if $\tilde(f)$ $k+1$, 
the real weight is $w_2$. From this analysis, we can know that the method
based on quantum counting is less efficient than our method because
the quantum counting method needs to exploit some period, which
requires several Grover operations.

\subsubsection{Find Real Weight $w$ from $n$ Possible Weights}
We can extend the previous result to a more general case. When $n$
possible weights are given. Can we decide which weight is real one?
The most important thing in this problem is to find the smallest
integer value of $P_s = LCM(a_1, a_2, \cdots, a_n)$. Then, as the same
method in the previous section, $k_i$ should be $P_s/a_i$. From this
analysis, we can find the real weight from the measured value
$\tilde{f}$ as if $\tilde{f}$ is $k_i$ then the real weight is
$w_i$. In this case, the time complexity is $P_s = LCM(a_1, a_2,
\cdots, a_n)$.

\section{Conclusion and Open Problems}
In this work, we have investigated the application of Grover
operators to distinguish weight of Boolean function when two weights
are given. Firstly, when we assume that the weight is the number of
solutions, we found that Algorithm
\ref{algo:randomized_weight_decision_algorithm} can find the exact
weight with almost certainty. Secondly, by exploiting the sure
success Grover search method, we found a sure success weight
decision algorithm, Algorithm
\ref{algo:sure_success_weight_decision_algorithm}, with modification
of the last two Grover operations with phase conditions. Lastly, we
have compared the proposed method to the quantum counting algorithm.
Because the quantum counting algorithm needs period information,
which requires more Grover operations, our method is more efficient
than the quantum counting method.

On the other hand, until this work, we assume that two weights are
given such as $wN$ and $(1-w)N$. How about other cases such as $w_1$
and $w_2$, where $w_1 + w_2 \neq 1$? Moreover, when three or more
weights are given, can we find the exact weight with the similar
approach? This may be attempted using the brief idea presented in
Subsection~\ref{difwt}.

\ \\
\noindent{\bf Acknowledgment:} Samuel L. Braunstein currently holds a Royal
Society - Wolfson Research Merit Award. Byung-Soo Choi is supported by the
Ministry of Information \& Communication of Korea
(IT National Scholarship Program).

\bibliographystyle{plain}

\begin{thebibliography}{10}

\bibitem{LA93}
Laurent Alonso, Edward M. Reingold, Rene Schott.
\newblock Determining the Majority.
\newblock {\em Information Processing Letters}, 47(5):253-255 (1993).

\bibitem{2000quant.ph..5055B}
G.~{Brassard}, P.~{Hoyer}, M.~{Mosca}, and A.~{Tapp}.
\newblock {Quantum Amplitude Amplification and Estimation}.
\newblock {\em quant-ph/0005055}, May 2000.

\bibitem{1998quant.ph..5082B}
G.~{Brassard}, P.~{Hoyer}, and A.~{Tapp}.
\newblock {Quantum Counting}.
\newblock {\em quant-ph/99805082}, May 1998.

\bibitem{BO98}
M.\ Boyer, G.\ Brassard, P.\ Hoyer and A.\ Tapp.
\newblock Tight Bounds on Quantum Searching.
\newblock Fortschr. Phys. 46 (1998) 4--5, 493--505.

\bibitem{CS-0900.81019MA}
D.~Deutsch.
\newblock {Quantum theory, the Church-Turing principle and the universal
  quantum computer.}
\newblock {\em Proc. R. Soc. Lond., Ser. A}, 400(1818):97--117, 1985.

\bibitem{qDJ92}
D.~Deutsch and R.~Jozsa.
\newblock {Rapid solution of problems by quantum computation.}
\newblock {\em Proc. R. Soc. Lond., Ser. A}, 439(1907):553--558, 1992.

\bibitem{GP01}
F.~Green and R.~Pruim.
\newblock Relativized Separation of EQP from ${\mbox P}^{\mbox{NP}}$.
\newblock {\em Information Processing Letters}, 80(5):257--260, 2001.

\bibitem{qGR96}
L.~Grover.
\newblock A fast quantum mechanical algorithm for database search.
\newblock In {\em Proceedings of 28th Annual Symposium on the Theory of
Computing (STOC)}, May 1996, pages 212--219. Available at
xxx.lanl.gov/quant-ph/9605043.

\bibitem{HOYER}
P.~H{\o}yer.
\newblock {Arbitary Phases in Quantum Amplitude Amplification}.
\newblock {\em Phys. Rev. A}, 62(5):052304, 2000.

\bibitem{LI-HWANG-HSIEH-WANG}
Jin-Yuan Hsieh, Kuo-Shong~Wang, Che-Ming~Li, Chi-Chuan~Hwang.
\newblock {General Phase-Matching Condition for a Quantum Searching Algorithm}.
\newblock {\em Phys. Rev. A}, 65(3):034305, 2002.

\bibitem{HSIEH-LI}
Che-Ming~Li, Jin-Yuan~Hsieh.
\newblock {General SU(2) Formulation for Quantum Searching with Certainty}.
\newblock {\em Phys. Rev. A}, 65(5):052322, 2002.

\bibitem{LONG-TU-LI-ZHANG-YAN}
Yan Song Li, Wei Lin, Zhang Hai, Yang~Yan, Gui Lu~Long, Chang Cun~Tu.
\newblock {An SO(3) Picture for Quantum Searching}.
\newblock {\em J. Phys. A, Math. Gen.}, 34:861, 2000.

\bibitem{LONG}
G.~L.~Long.
\newblock {Grover Algorithm with Zero Theoretical Failure Rate}.
\newblock {\em Phys. Rev. A}, 64(2):022307, 2001.

\bibitem{MM01}
Michele Mosca.
\newblock Counting by quantum eigenvalue estimation.
\newblock {\em Theoretical Computer Science}, 264(1):139--153, 2001.

\bibitem{AN98}
Ashwin Nayak, Felix Wu.
\newblock The quantum query complexity of approximating median and related
statistics.
\newblock Available at xxx.lanl.gov/quant-ph/9804066.

\bibitem{qNC02}
M.\ A.\ Nielsen and I.\ L.\ Chuang.
\newblock Quantum Computation and Quantum Information.
\newblock Cambridge University Press, 2002.

\bibitem{SICOMP::Shor1997}
P.~W.~Shor.
\newblock Polynomial-time algorithms for prime factorization and discrete
  logarithms on a quantum computer.
\newblock {\em SIAM Journal on Computing}, 26(5):1484--1509, October 1997.

\bibitem{LONG-LI-SUN}
Yang~Sun, Gui-Lu~Long, Xiao~Li.
\newblock {Phase Matching Condition for Quantum Search with a Generalized
  Initial State}.
\newblock {\em Phys. Let. A}, 294:143, 2002.

\end{thebibliography}

\end{document}